\documentclass[final,5p,times,twocolumn]{elsarticle}
\pdfoutput=1

\usepackage{lineno}
\usepackage{atlasphysics}
\usepackage{subfigure}
\usepackage{multirow}
\usepackage{flafter}

\biboptions{numbers,square,comma,sort&compress}

\usepackage{dblfloatfix}
\usepackage{fixltx2e}
\usepackage[hyperindex,breaklinks]{hyperref} 
\usepackage{url}
\usepackage{placeins}

\newcommand\sevenTeV{\ensuremath{7\tev}}
\newcommand\eightTeV{\ensuremath{8\tev}}
\newcommand\sqrtsevenTeV{\ensuremath{\sqrt{s} = 7\tev}}
\newcommand\sqrteightTeV{\ensuremath{\sqrt{s} = 8\tev}}
\renewcommand\mH{\ensuremath{m_{\Hboson}}}
\newcommand\hgg{\ensuremath{\Hboson \rightarrow \gamma\gamma}}

\newcommand\ttH{\ensuremath{\ttbar\Hboson}}

\newcommand\tH{\ensuremath{t\Hboson}}
\newcommand\tHqb{\ensuremath{t\Hboson{q}{b}}}
\newcommand\WtH{\ensuremath{Wt\Hboson}}

\newcommand\tbarH{\ensuremath{\bar{t}\Hboson}}
\newcommand\ggH{ggF}
\newcommand\VBFH{VBF}
\newcommand\WH{\ensuremath{W\Hboson}}
\newcommand\ZH{\ensuremath{Z\Hboson}}

\newcommand{\mgamgam}{\ensuremath{m_{\gamma\gamma}}} 
\newcommand\mgg{\mgamgam}

\newcommand\muttH{\ensuremath{\mu_{\ttH}}}
\newcommand\muH{\ensuremath{\mu}}
\newcommand\mHcomb{125.4}

\newcommand\abstracttextnew{
  A search is performed for Higgs bosons produced in association with top quarks using
  the diphoton decay mode of the Higgs boson. Selection requirements are optimized separately for leptonic and fully
  hadronic final states from the top quark decays.
  The
  dataset used corresponds to an integrated luminosity of 4.5~\ifb\
  of proton--proton collisions at a center-of-mass energy of \sevenTeV\ and 20.3~\ifb\ at \eightTeV\
  recorded by the ATLAS detector at the CERN Large Hadron Collider. No
  significant excess over the background prediction is observed and upper limits
  are set on the \ttH\ production cross section.
  The observed exclusion upper limit at 95\% confidence level is
  6.7 times the predicted Standard Model cross section value.
  In addition, limits are set on the strength of the Yukawa coupling between the top quark and the Higgs boson,
  taking into account the dependence of the \ttH\ and \tH\ cross sections as well as the \hgg\ branching fraction on the Yukawa coupling.
  Lower and upper limits at 95\% confidence level are set at $-$1.3 and +8.0 times the Yukawa coupling strength in the Standard Model.
}

\newcommand\titletext{Search for \hgg\ produced in association with top quarks and constraints on the Yukawa coupling between the top quark and the Higgs boson using data taken at \sevenTeV\ and \eightTeV\ with the ATLAS detector}

\usepackage{preprintcover}
\PreprintCoverPaperTitle{Search for \hgg\ produced in association with top quarks and constraints on the Yukawa coupling between the top quark and the Higgs boson using data taken at \sevenTeV\ and \eightTeV\ with the ATLAS detector}
\PreprintIdNumber{CERN-PH-EP-2014-179}  
\PreprintCoverAbstract{  
  A search is performed for Higgs bosons produced in association with top quarks using
  the diphoton decay mode of the Higgs boson. Selection requirements are optimized separately for leptonic and fully
  hadronic final states from the top quark decays.
  The
  dataset used corresponds to an integrated luminosity of 4.5~\ifb\
  of proton--proton collisions at a center-of-mass energy of \sevenTeV\ and 20.3~\ifb\ at \eightTeV\
  recorded by the ATLAS detector at the CERN Large Hadron Collider. No
  significant excess over the background prediction is observed and upper limits
  are set on the \ttH\ production cross section.
  The observed exclusion upper limit at 95\% confidence level is
  6.7 times the predicted Standard Model cross section value.
  In addition, limits are set on the strength of the Yukawa coupling between the top quark and the Higgs boson,
  taking into account the dependence of the \ttH\ and \tH\ cross sections as well as the \hgg\ branching fraction on the Yukawa coupling.
  Lower and upper limits at 95\% confidence level are set at $-$1.3 and +8.0 times the Yukawa coupling strength in the Standard Model.
}
\PreprintJournalName{PLB}

\begin{document}
\begin{frontmatter}
\title{\titletext}
\author{The ATLAS Collaboration}
\address{}

\begin{abstract}
\abstracttextnew
\end{abstract}
\end{frontmatter}

\section{Introduction}

After the decades-long search for the Higgs boson~\cite{Englert:1964et,Higgs:1964pj,higgs3},
a particle consistent with the Standard Model (SM) Higgs boson has been discovered at the Large Hadron Collider (LHC)~\cite{2012gu,2012gk}.
A notable property of the SM Higgs boson is its predicted large Yukawa coupling to top quarks, $Y_t^{\rm SM}$.
The measurement of $Y_t$ is particularly important for understanding electroweak symmetry breaking and allows for testing theories beyond the SM (BSM).

The value of $Y_t$ is indirectly tested by measurements sensitive to gluon fusion, \ggH,
the dominant Higgs boson production mechanism at the LHC, which receives large contributions from loop diagrams involving the top quark.
In addition, $Y_t$ is probed in the decay of the Higgs boson to two photons, \hgg, 
as the decay width also involves loop diagrams with top quarks~\cite{Aad:2013wqa}.
However, $Y_t$ can be directly measured in the production of top--antitop quark pairs, \ttbar, in association with a Higgs
boson~\cite{Beenakker:2001rj,Reina:2001bc,Dawson:2003zu,Garzelli:2011vp,Frederix:2011zi}, \ttH.

\begin{figure}[!h]
\centering
\subfigure[]{
  \includegraphics[width=0.2\textwidth]{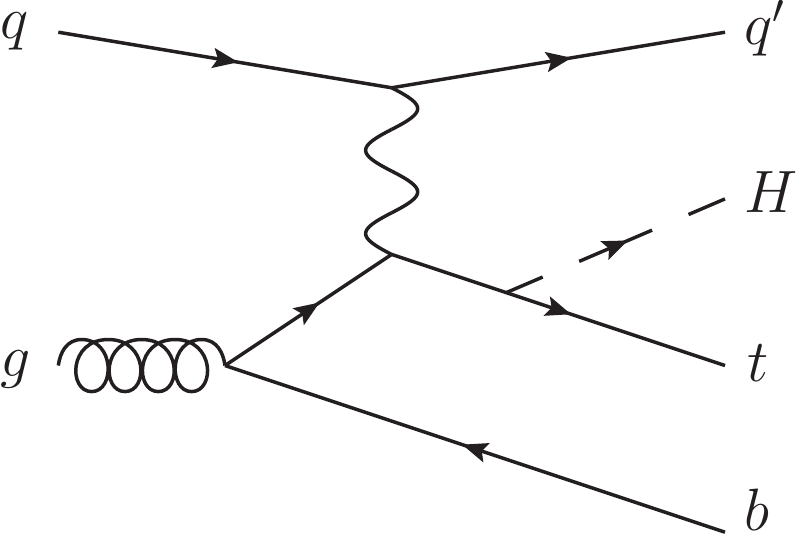}
}
\subfigure[]{
  \includegraphics[width=0.2\textwidth]{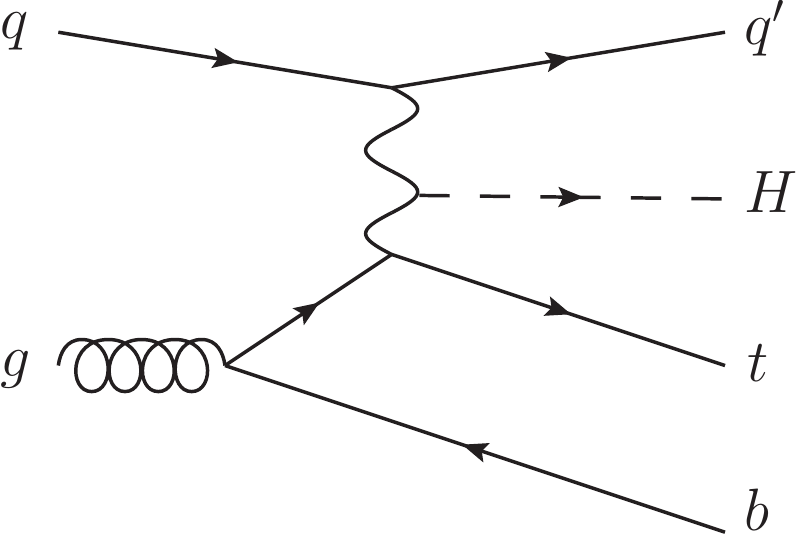}
}
\subfigure[]{
  \includegraphics[width=0.2\textwidth]{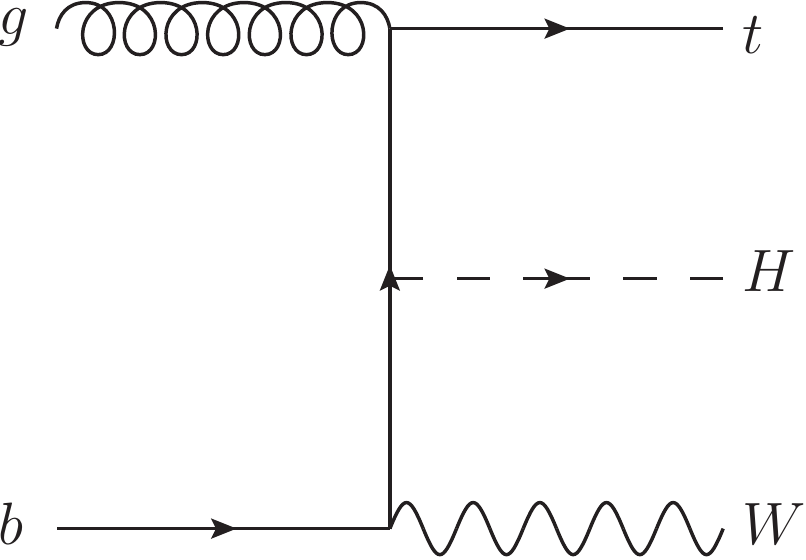}
}
\subfigure[]{
  \includegraphics[width=0.2\textwidth]{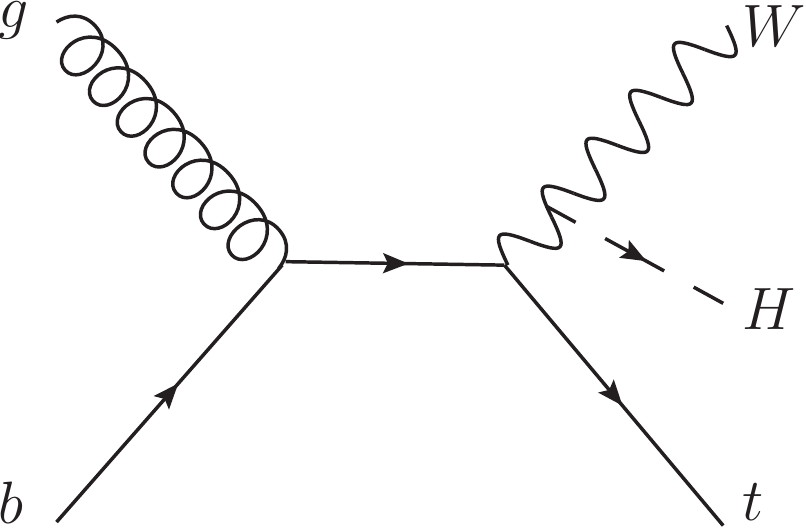}
}
\caption{Feynman diagrams showing examples for \tHqb\ (a, b) and \WtH\ production (c, d). Higgs boson radiation off top quark and $W$ boson lines
is depicted. The \tHqb\ process is shown in the four-flavor scheme where no $b$-quarks are assumed to be present in the proton~\cite{Frederix:2012dh}.
}
\label{fig:feynman}
\end{figure}

The production of the Higgs boson in association with a single top quark, \tH,\footnote{
For simplicity, \tH\ refers equally to \tbarH\ in this Letter.
}
is also sensitive to $Y_t$.
Three processes contribute to \tH\ production~\cite{Maltoni:2001hu,Barger:2009ky,Farina:2012xp,Biswas:2012bd,Agrawal:2012ga}:
$t$-channel (\tHqb) production,
\WtH\ production and $s$-channel \tH\ production.
The $s$-channel production is neglected in this Letter due to the much smaller cross section compared to \tHqb\ and \WtH\ production.
Examples of Feynman diagrams for \tHqb\ and \WtH\ production are shown in Fig.~\ref{fig:feynman}.

In the SM, \tH\ production is suppressed by the destructive interference between $t$-channel diagrams with Higgs bosons emitted from top quark and $W$ boson lines,
as for example shown in Fig.~\ref{fig:feynman} (a) and Fig.~\ref{fig:feynman} (b).
In BSM theories~\cite{Barger:2009ky,Farina:2012xp,Biswas:2012bd,Agrawal:2012ga}, however, $Y_t$ can have non-SM values, and in particular
the relative sign between $Y_t$ and $g_{HWW}$, which quantifies the coupling between the Higgs boson and the $W$ boson, can be different from the SM prediction, 
which could lead to constructive instead of destructive interference in \tH\ production.
Hence, the \tH\ production cross section is not only sensitive to the magnitude of $Y_t$ but, in contrast to \ttH\ production, it is also sensitive to the relative sign of
$Y_t$ with respect to $g_{HWW}$.
A scale factor, $\kappa_t$, is introduced to describe the relation between $Y_t$ and its SM value: $Y_t = \kappa_t Y_t^{\rm SM}$.
Values of $\kappa_t \neq 1$ imply modifications of the Brout--Englert--Higgs mechanism and are assumed here to leave the top quark mass and decay properties unchanged.
Furthermore, only SM particles are assumed to contribute to the decay width of the Higgs boson.

This Letter reports a search for \hgg\ in association with top quarks using data recorded with the ATLAS detector~\cite{detectorpaper}.
Measurements in the \hgg\ decay channel are challenging due to the small branching fraction in the SM, ${\rm{BR(}}\hgg{\rm{)}} = 2.28 \times 10^{-3}$ for Higgs boson masses, \mH, around 125~\gev.
However, the diphoton final state allows the diphoton invariant mass, \mgg, to be reconstructed with excellent resolution, strongly reducing the contribution from
the backgrounds, which have a falling \mgg\ spectrum,
referred to as continuum background in the following.
The contribution from the continuum background can be derived from data sidebands, thus not relying on theory assumptions.
A previous search for \ttH\ production by the CMS Collaboration has explored hadronic, diphoton and leptonic final states of the
Higgs boson~\cite{Khachatryan:2014qaa}, setting an upper limit at the 95\% confidence level (CL) on the ratio of the observed \ttH\ production cross section to the SM expectation, called the signal strength \muttH, of 4.5.

This Letter also reports lower and upper limits at 95\% CL on $\kappa_t$, taking into account the changes in the \ttH\ and \tH\ cross sections as well as the
\hgg\ branching fraction~\cite{Farina:2012xp,Biswas:2012bd,Agrawal:2012ga}.
BSM theories with values of $Y_t \neq Y_t^{\rm{SM}}$ are hence constrained.

\section{The ATLAS detector}

The ATLAS detector consists of an inner tracking detector system,
electromagnetic and hadronic calorimeters, and an external muon spectrometer. 
Charged particles in the pseudorapidity
\footnote{ATLAS uses a right-handed coordinate system with its origin at the nominal interaction point (IP) in the centre of the detector and the $z$-axis along the beam pipe. The $x$-axis points from the IP to the centre of the LHC ring, and the $y$-axis points upward. Cylindrical coordinates $(r,\phi)$ are used in the transverse plane, $\phi$ being the azimuthal angle around the beam pipe. The pseudorapidity is defined in terms of the polar angle $\theta$ as $\eta=-\ln\tan(\theta/2)$.
The transverse momentum is defined as $\pt=p\sin\theta=p/\cosh\eta$,
and the transverse energy $\et$ has an analogous definition.}
range \mbox{$|\eta| < 2.5$} are reconstructed with the inner tracking detector, which is immersed in a \mbox{$2\ \mathrm{T}$} axial field
provided by a superconducting solenoid, and consists of pixel and microstrip semiconductor detectors, as well as a straw-tube transition radiation tracker.
The solenoid is surrounded by sampling calorimeters, which span the pseudorapidity range up to \mbox{$|\eta|$ = 4.9}.
High-granularity liquid-argon (LAr) electromagnetic calorimeters are present up to \mbox{$|\eta|$ = 3.2}.
Hadronic calorimeters with scintillator tiles as active material cover \mbox{$|\eta| <$ 1.74}, while LAr technology is used for hadronic calorimetry from
\mbox{$|\eta|$ = 1.5} to \mbox{$|\eta|$ = 4.9}.
Outside the calorimeter system, air-core toroids provide a magnetic field for the muon spectrometer. 
Three stations of precision drift tubes and cathode strip chambers provide a measurements of muon tracks in the region
\mbox{$|\eta| < 2.7$}.
Resistive-plate and thin-gap chambers provide muon triggering capability up to \mbox{$|\eta| < 2.4$}.
A detailed description of the ATLAS detector can be found in Ref.~\cite{detectorpaper}.

\section{Data and Monte Carlo samples}

\subsection{Data samples}

Data used for this analysis were recorded in $pp$ collisions at \sqrtsevenTeV\ and \eightTeV\ in 2011 and 2012, respectively.
All events satisfy data quality requirements ensuring proper functioning of the detector and trigger subsystems. The
resulting datasets correspond to integrated luminosities of 4.5~\ifb\ and 20.3~\ifb, respectively~\cite{lumi}.
For the \sevenTeV\ dataset, events were triggered with a diphoton trigger with a threshold of 20~\gev\ on the transverse energy of each
photon candidate.
For the \eightTeV\ dataset, these thresholds were raised to 35~\gev\ for the highest-\et\ (leading) photon candidate and
25~\gev\ for the second-highest-\et\ (subleading) photon candidate.

\subsection{Monte Carlo samples}

The contribution from the continuum background is directly estimated from data.
All processes involving \hgg\ decays, however, are estimated using Monte Carlo (MC) simulation samples.

The production of \ttH\ events is modeled using next-to-leading-order (NLO) matrix elements obtained with the {\sc HELAC}-Oneloop
package~\cite{Bevilacqua:2011xh}, where {\sc Powheg-BOX}~\cite{Nason:2004rx,Frixione:2007vw,Alioli:2010xd} is interfaced
to {\sc Pythia} 8.1~\cite{Pythia8} for showering and hadronization.
{\sc CT10}~\cite{Lai:2010vv} parton distribution functions (PDF) and the {\sc AU2} underlying event
tune~\cite{ATLAS:2011gmi,ATLAS:2011zja} are used.
Production of \tHqb\ is simulated with
{\sc MadGraph}~\cite{Maltoni:2002qb}
in the four-flavor scheme with the CT10 PDF set,
which provides a better description of the kinematics of the spectator $b$-quark than the five-flavor scheme~\cite{Frederix:2012dh}.
{\sc Pythia} 8.1 is used for showering and hadronization.
Production of \WtH\ is simulated in the five-flavor scheme by {\sc MadGraph5\_aMC@NLO}~\cite{Alwall:2014hca} interfaced to Herwig++~\cite{Bahr:2008pv} using the CT10 PDF set.
All \tH\ samples are produced for three different values of $\kappa_t$: $-1$, 0 and +1.
In the simulation of \ttH, \tHqb\ and \WtH\ processes, diagrams with Higgs bosons radiated in the top quark decay are not taken into account
because such contributions are negligible~\cite{Han:2013sea}.

Higgs boson production by \ggH\ and vector-boson fusion (\VBFH) is simulated with {\sc Powheg-BOX}~\cite{powheg_ggF, Nason:2009ai}
interfaced to {\sc Pythia} 8.1 for showering and hadronization with CT10 PDF.
Production of a Higgs boson in association with a $W$ or $Z$ boson (\WH, \ZH) is simulated with {\sc Pythia} 8.1
using CTEQ6L1~\cite{Nadolsky:2008zw} PDF.

All MC samples are generated at $\mH = 125\gev$ and
are passed through a full GEANT4~\cite{geant4} simulation of the ATLAS detector~\cite{atlasSim}.
The simulated samples have additional $pp$ collision events, pile-up, simulated by {\sc Pythia} 8.1 added and weighted such that the average number
of interactions per bunch-crossing is the same as in data.

The cross sections for \ttH\ production were calculated at NLO in
quantum chromodynamics (QCD)~\cite{Beenakker:2001rj,Beenakker:2002nc,Dawson:2002tg,Dawson:2003zu}.
The cross sections for \tHqb\ production are calculated for different values of $\kappa_t$ at LO
using {\sc MadGraph} with the renormalization and factorization scales set to 75~\gev,
and with a minimum $p_{\mathrm{T}, q}$ requirement of 10~\gev, consistent with the generated MC samples.
LO-to-NLO K-factors are obtained by comparing the LO cross sections with the NLO cross sections calculated using {\sc MadGraph5\_aMC@NLO}.
The cross sections for \WtH\ production are calculated for different values of $\kappa_t$ at NLO using {\sc MadGraph5\_aMC@NLO} with
dynamic renormalization and factorization scales. Interference effects with \ttH\ production are not considered, but are believed to
be small given that \WtH\ is produced mostly without a second high-\pt\ $b$-quark in the final state.

The cross sections for \ggH\ production were calculated at next-to-next-to-leading order (NNLO) in QCD~\cite{Djouadi:1991tka,Dawson:1990zj,Spira:1995rr,Harlander:2002wh,Anastasiou:2002yz,Ravindran:2003um}. In addition, QCD soft-gluon
resummation up to next-to-next-to-leading logarithms~\cite{Catani:2003zt} is adopted to improve the NNLO calculation, and NLO electroweak
(EW) corrections are applied~\cite{Aglietti:2004nj,Actis:2008ug}.
The cross sections for \VBFH\ production were calculated including NLO QCD and EW
corrections~\cite{Ciccolini:2007jr,Ciccolini:2007ec,Arnold:2008rz}.
In addition, approximate NNLO QCD corrections are applied~\cite{Bolzoni:2010xr}.
The cross sections for \WH\ and \ZH\ production were calculated at NLO~\cite{Han:1991ia}
and NNLO~\cite{Brein:2003wg} in QCD.
Moreover, NLO EW corrections~\cite{Ciccolini:2003jy} are applied.

The theoretical uncertainties on the Higgs boson production cross sections come from varying the renormalization and
factorization scales and from uncertainties on the parton distribution functions ~\cite{Botje:2011sn,Lai:2010vv,Martin:2009iq,Ball:2011mu}. 
The Higgs boson decay branching fractions are taken from Refs.~\cite{Djouadi:1997yw,Bredenstein:2006rh, Actis:2008ts, Denner:2011mq} and
their uncertainties are compiled in Refs.~\cite{LHCHiggsCrossSectionWorkingGroup:2011ti, LHCHiggsCrossSectionWorkingGroup:2012vm}.
A summary of the cross-section values and their uncertainties is given in Table~\ref{tab:xsecs}.

\renewcommand{\arraystretch}{1.2}
\begin{table}[!h]
  \centering
  \caption{Production cross sections for the various Higgs boson processes at \sevenTeV\ and \eightTeV\ before taking into account the BR(\hgg) at $\mH = 125~\gev$.
    Also quoted are the theoretical uncertainties from variations of the renormalization and factorization scales and uncertainties
    on the parton distribution functions~\cite{LHCHiggsCrossSectionWorkingGroup:2011ti, LHCHiggsCrossSectionWorkingGroup:2012vm}.
  }
  \vspace{0.25cm}
  \begin{tabular}{l|c|c}
    Process & $\sigma$ [pb] at \sevenTeV & $\sigma$ [pb] at \eightTeV \\
    \hline
    \ttH                   & $0.086^{+0.008}_{-0.011}$    & $0.129^{+0.012}_{-0.016}$      \\
    \tHqb, $\kappa_t = +1$ & $0.0111^{+0.0009}_{-0.0008}$ & $0.0172^{+0.0012}_{-0.0011}$    \\
    \tHqb, $\kappa_t = 0$  & $0.040^{+0.003}_{-0.003}$    & $0.059^{+0.004}_{-0.004}$      \\
    \tHqb, $\kappa_t = -1$ & $0.129^{+0.010}_{-0.009}$    & $0.197^{+0.014}_{-0.013}$        \\
    \WtH, $\kappa_t = +1$  & $0.0029^{+0.0007}_{-0.0006}$ & $0.0047^{+0.0010}_{-0.0009}$     \\
    \WtH, $\kappa_t = 0$   & $0.0043^{+0.0011}_{-0.0008}$ & $0.0073^{+0.0017}_{-0.0013}$     \\
    \WtH, $\kappa_t = -1$  & $0.016^{+0.004}_{-0.003}$    & $0.027^{+0.006}_{-0.005}$       \\
    ggF                    & $15.1 \pm 1.6$               & $19.3\pm 2.0$               \\
    VBF                    & $1.22\pm 0.03$               & $1.58\pm 0.04$              \\
    \WH                    & $0.579\pm 0.016$             & $0.705\pm 0.018$            \\
    \ZH                    & $0.335\pm 0.013$             & $0.415\pm 0.017$            \\
  \end{tabular}
  \label{tab:xsecs}
\end{table}
\renewcommand{\arraystretch}{1.0}

\section{Object and event selection}

\begin{table*}[t]
  \centering
  \caption{
    Expected numbers of \hgg\ events ($N_H$) from an SM Higgs boson with $\mH = \mHcomb\gev$ after the event selection.
    These combined yields are normalized to 4.5~\ifb\ for the \sevenTeV\ data and to 20.3~\ifb\ for the \eightTeV\ data, and are listed in the table along with the percent contribution of each Higgs boson production process with respect to the sum of all Higgs boson production processes.
    The numbers of fitted continuum background events~($N_B$) for the \sevenTeV\ and \eightTeV\ data are also shown, 
    where $N_B$ is the integral of the continuum background in the \mgg\ range 120--130~\GeV\ , 
    which is determined by an unbinned signal-plus-background fit to all categories with one common scale factor for the \hgg\ normalization.
    The uncertainty on $N_B$ is the statistical uncertainty calculated from $\delta{N_{B}}=\delta{N_{\rm{tot}}}N_{B}/N_{\rm{tot}}$, 
    where $N_{\rm{tot}}$ is the total number of background events in the full \mgg\ range 105--160~\gev\ estimated from an unbinned signal-plus-background likelihood fit, and $\delta N$ denotes the Poisson uncertainty on $N$.
  }
  \vspace{0.25cm}

  \begin{tabular}{lc|ccccccc|c}
    Category & $N_H$  & \ggH   & \VBFH  & \WH  & \ZH  & \ttH  & \tHqb  & \WtH  & $N_{B}$ \\
    \hline  
   \sevenTeV\ leptonic selection &     0.10     &   0.6   & 0.1   & 14.9  & 4.0   & 72.6          & 5.3   & 2.5          & $ 0.5^{+0.5}_{-0.3}$\\
   \sevenTeV\ hadronic selection &     0.07     &   10.5  & 1.3   & 1.3   & 1.4   & 80.9          & 2.6   & 1.9          & $ 0.5^{+0.5}_{-0.3}$\\
   \eightTeV\ leptonic selection &     0.58     &   1.0   & 0.2   & 8.1   & 2.3   & 80.3          & 5.6   & 2.6          & $ 0.9^{+0.6}_{-0.4}$\\
   \eightTeV\ hadronic selection &     0.49     &   7.3   & 1.0   & 0.7   & 1.3   & 84.2          & 3.4   & 2.1          & $ 2.7^{+0.9}_{-0.7}$\\
  \end{tabular}
  \label{tab:yields}
\end{table*}

\begin{figure*}[t]
\centering
\subfigure[]{
  \includegraphics[width=0.47\textwidth]{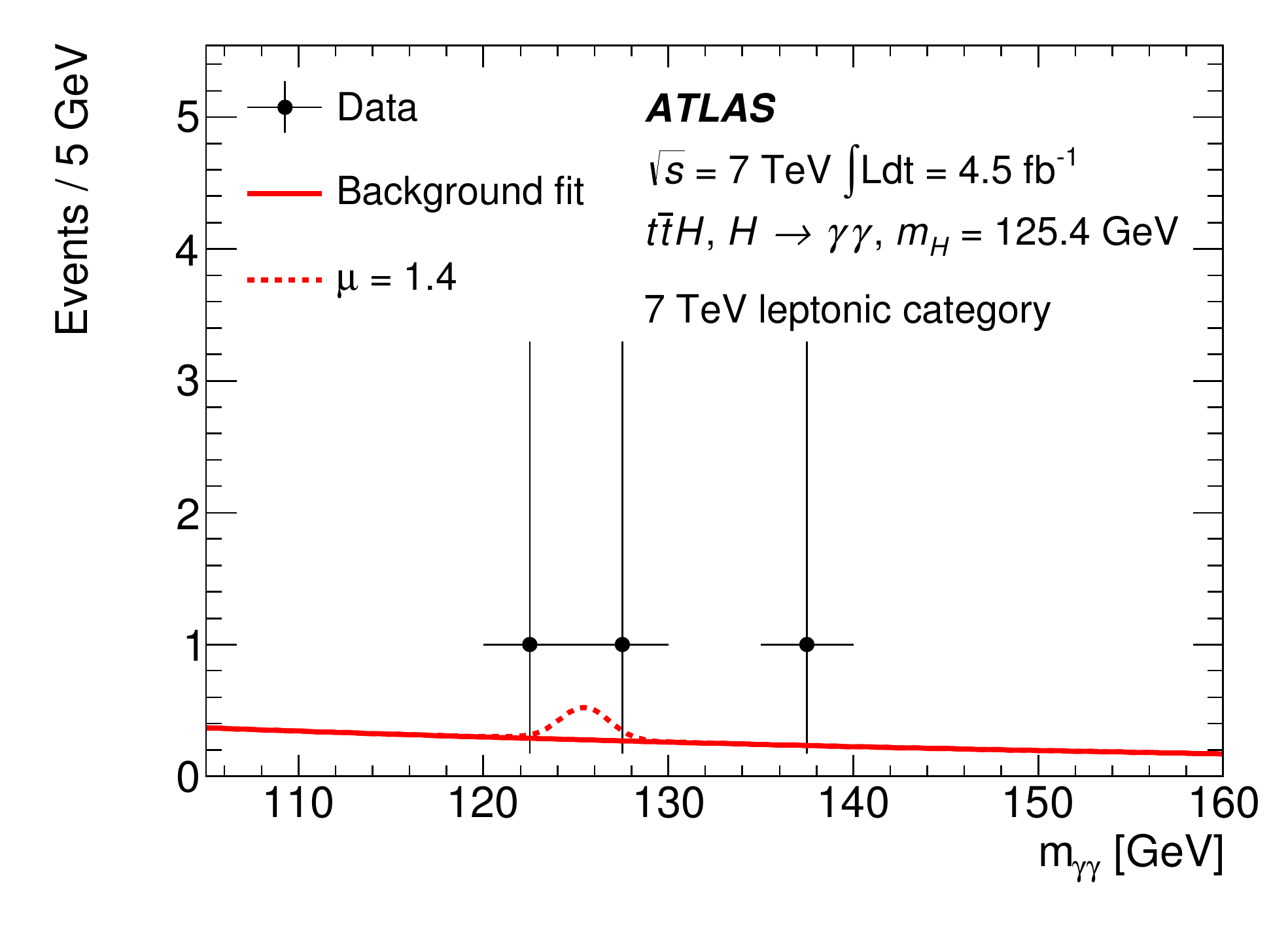}
}
\subfigure[]{
  \includegraphics[width=0.47\textwidth]{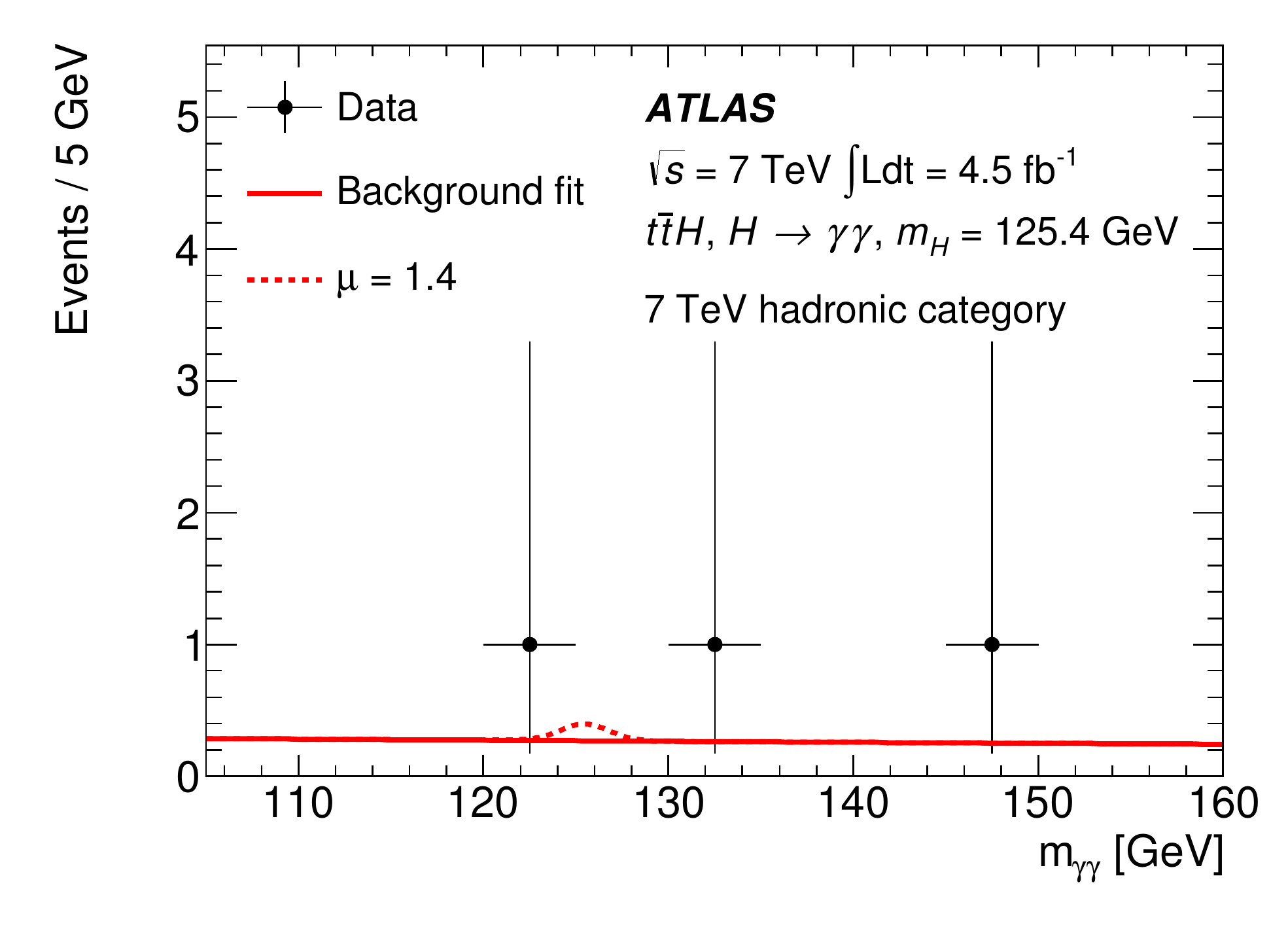}
}
\subfigure[]{
  \includegraphics[width=0.47\textwidth]{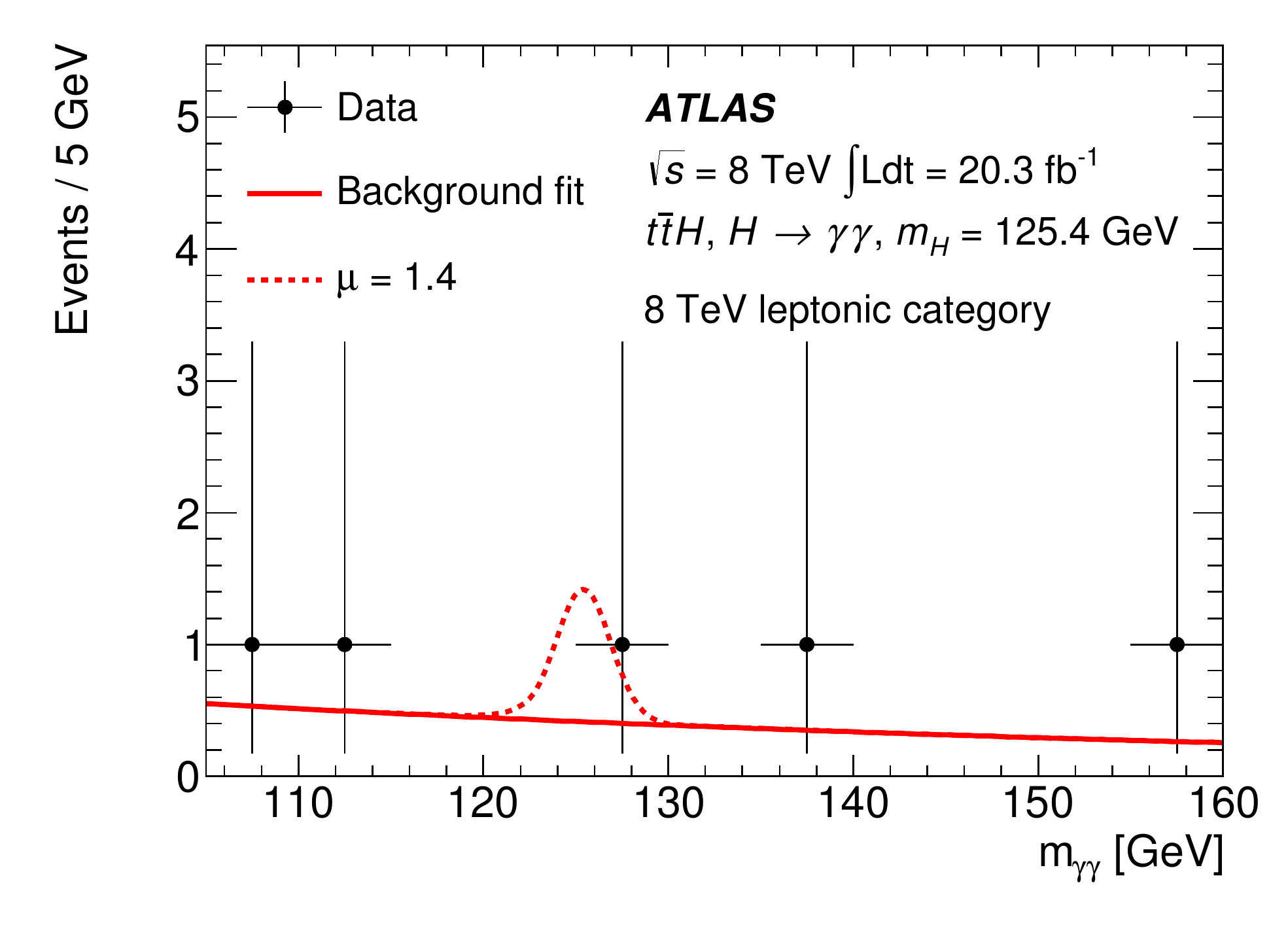}
}
\subfigure[]{
  \includegraphics[width=0.47\textwidth]{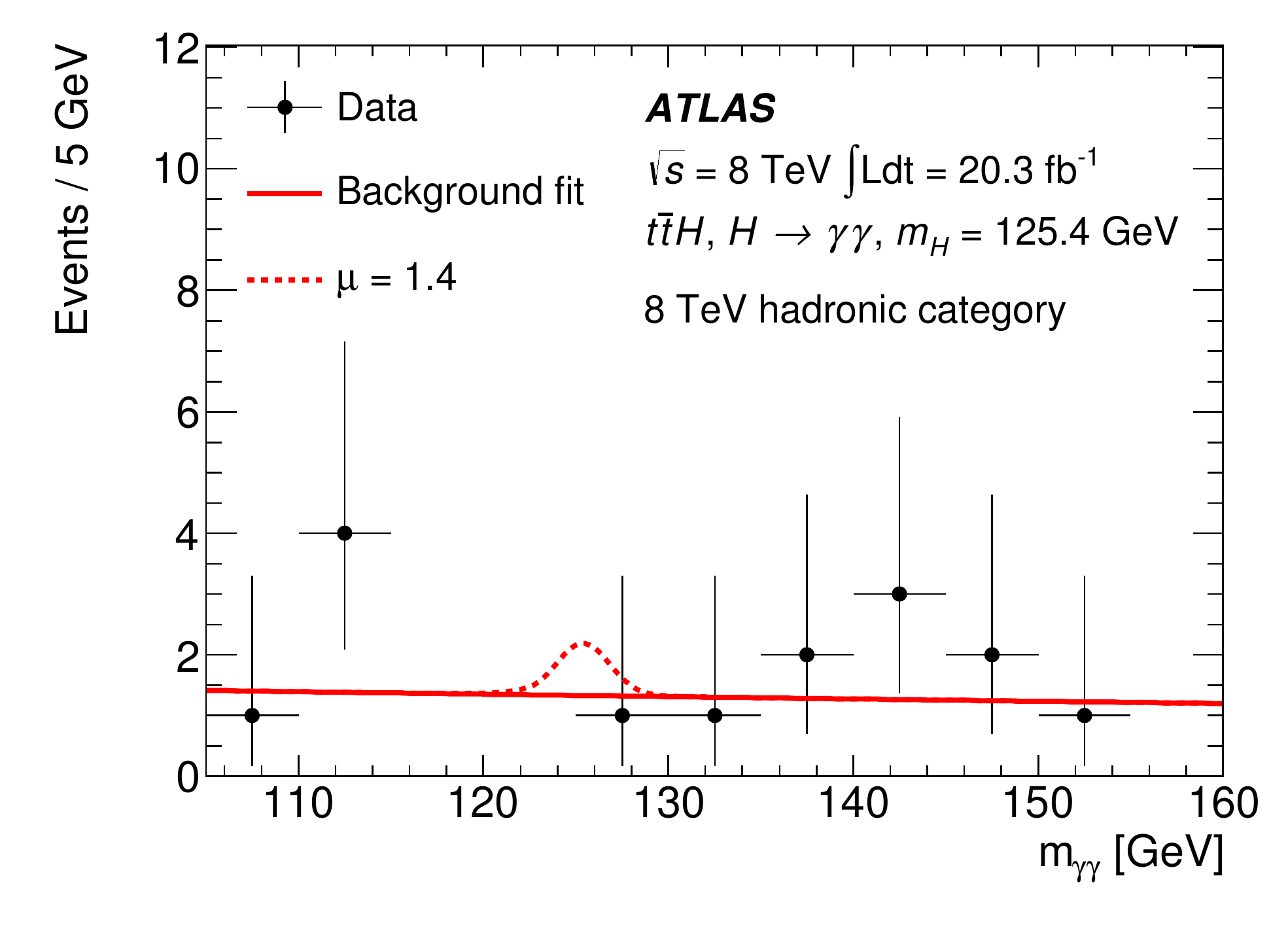}
}

\caption{Distributions of the diphoton invariant mass, \mgg, for the leptonic (left) and hadronic (right) category for data at \sevenTeV\ (top) and data at \eightTeV\ (bottom).
  An unbinned signal-plus-background likelihood fit to the full spectra is used to
  estimate the number of events from continuum background (solid line) as well as from SM Higgs boson production (dashed line).
  The signal strength, $\mu$, is a parameter common to all categories and its best-fit value is $\mu = 1.4$ for $\mH = \mHcomb\gev$.
}
\label{fig:fit}
\end{figure*}

\subsection{Object selection}
\label{sec:objandevent:obj}

Photons are reconstructed~\cite{PhotonPerformance} from clusters of cells in the electromagnetic calorimeter in the region $|\eta| < 2.37$ excluding the transition region,
$1.37 < |\eta| < 1.56$, between the barrel and endcap calorimeters.
Unconverted photons are required to have no tracks associated with them; 
clusters from photons converted in the material between the production vertex and the calorimeter are allowed to have one or two
associated tracks.
The energies of the clusters are calibrated, separately for unconverted and converted photon candidates, in order to account for energy losses
upstream of the calorimeter and for energy leakage outside of the cluster.
Photons are required to pass a set of selection requirements on the reconstructed shower shape as well as
the following isolation requirements:
the sum of the \pt\ of all particles featuring tracks with $\pt > 1 \GeV$ 
in a cone of size
$\Delta R \equiv \sqrt{\left(\Delta\eta\right)^2+\left(\Delta\phi\right)^2} = 0.2$ around the photon is required to be smaller than 2.6 (2.2) \GeV\ for the
\sqrteightTeV\ (\sevenTeV) data.
Tracks from converted photons are excluded from the sum.
Moreover, the sum of the $\et$ values in the calorimeter cells in a cone of size $\Delta R = 0.4$ around the photon is required to be smaller than 6 (5.5) \GeV\
for the \eightTeV\ (\sevenTeV) data.
The calorimeter isolation is corrected for photon energy leakage.
It is also corrected event-by-event by using the ambient energy from pile-up and the underlying event~\cite{Cacciari:2009dp,Aad:2010sp}.
Only events with two photons are retained and a diphoton vertex is reconstructed by a neural-network-based algorithm~\cite{MassPaper},
which uses as input the
trajectories of the two photons and the tracks associated with different vertex candidates.
The photon trajectory is determined from the longitudinal profile of the photon shower in the calorimeter, the average $pp$ collision point, and for converted photons from the
direction of the associated tracks.
The leading (subleading) photon is required to have $\et > $ $0.35\times\mgg$ ($0.25\times\mgg$),
and the diphoton mass is required to be between 105~\gev\ and 160~\gev.

Electrons are reconstructed~\cite{Aad:2014fxa} from clusters of cells in the electromagnetic calorimeter with an associated track.
Only clusters in the region $|\eta| < 2.47$ are considered and are required to fulfill requirements on their shape to be
consistent with an electron.
The electron $\et$ has to be larger than $15\GeV$.
In addition, electrons must be isolated: the $\et$ in a cone of size $\Delta R = 0.4$ around the electron
and the sum of the transverse momenta of the tracks in a cone of size $\Delta R = 0.2$ around the electron must be
smaller than 20\% and 15\% of the electron $\et$, respectively.

Muons are reconstructed~\cite{Aad:2014zya} by combining tracks in the inner detector with tracks or track-segments in the muon spectrometer.
Muons are required to satisfy $|\eta| < 2.7$ and $\pT > 10\GeV$ and have to be isolated:
muons closer than $\Delta R = 0.4$ to a jet or to one of the two photons are not considered.
Moreover, the $\et$ in a cone of size $\Delta R = 0.4$ around the muon
and the sum of the transverse momenta of the tracks in a cone of size $\Delta R = 0.2$ around the muon must be
smaller than 20\% and 15\% of the muon $\pt$, respectively.

Jets are reconstructed from clusters of cells in the calorimeter with the anti-$k_t$ algorithm~\cite{anti_kt} with a radius parameter of 0.4.
They are calibrated to the hadronic energy scale~\cite{Aad:2014bia}, and only those with $\pt > 25 \gev$ and $|\eta| < 2.5$ are considered.
The jet energy is corrected for energy deposits from additional soft interactions in the event~\cite{Cacciari2008119}.
In order to suppress jets from additional interactions, the jet vertex fraction (JVF) must be larger than 50\% for jets with $\pt < 50 \GeV$ and
$|\eta| < 2.4$. 
The JVF is defined from the summed track \pt\ as the fraction associated with the primary diphoton vertex, where all tracks with $\pt > 0.5 \GeV$ matched to the jet are considered.

Jets containing $b$-quarks are identified with a neural-network-based $b$-tagging algorithm, which combines variables from impact parameter, secondary vertex and decay
topology algorithms evaluating the track parameters associated with the jet~\cite{ATLAS-CONF-2011-102}.
Three different working points (WP) with efficiencies of 60\%, 70\% and 80\% for identifying $b$-jets are used for \eightTeV\ data.
For \sevenTeV\ data, a slightly different optimization of the $b$-tagging algorithm with a WP corresponding to an efficiency of 85\% is used.
The $b$-tagging and mistagging efficiencies are measured in data using dijet and $\ttbar$ events~\cite{ATLAS-CONF-2014-046}.

The magnitude of the missing transverse momentum in each event, \met, is calculated using clusters of cells in the calorimeter.
Corrections are applied for identified photons, electrons, muons and jets according to special \met\ object identification requirements~\cite{metperf}.

In order to avoid double-counting of reconstructed objects,
electrons with a distance in $\eta$--$\phi$ space smaller than 0.4 to one of the two photons, $\Delta R(e, \gamma)$, are not considered.
In addition, jets with $\Delta R({\rm jet},\gamma)<0.4$ or $\Delta R({\rm jet},e)<0.2$ are removed.

\subsection{Event selection}

In addition to the requirement of two good photons satisfying the criteria described in Section~\ref{sec:objandevent:obj}, two different event selections were optimized in order to efficiently select leptonic \ttH\ events (leptonic category)
as well as all-hadronic \ttH\ events (hadronic category).
The optimization targeted an optimal expected limit on the signal strength \muttH\ in case no evidence for \ttH\ production is found.
However, the requirements for the leptonic category are kept loose enough in order to also allow high selection efficiency for \tHqb\ and \WtH\ production.

In this analysis, we assume that the top quark only decays to a $W$ boson and a $b$-quark.
The leptonic selection targets both the single-lepton decays of the \ttbar\ pairs, where one of the $W$ bosons decays leptonically and the other one decays hadronically, and the dilepton decays of \ttbar\ pairs, where both $W$ bosons decay leptonically.
Events are selected by requiring at least one electron or muon, at least one $b$-tagged jet using the 80\% (85\%) WP for
\eightTeV\ (\sevenTeV) data and $\met > 20 \GeV$.
The \met\ requirement is imposed to reduce backgrounds from final states without top quarks and it is not used for events with two or more $b$-tagged jets.
Events with an electron--photon invariant mass in the range 84--94~\GeV\ are rejected in order to reduce the background contribution
from $Z\to ee$ events with one electron misidentified as a photon.

The hadronic selection targets events where both $W$ bosons, from the top quark decays, decay hadronically. No electrons or muons may be identified in the event.
Events must fulfill requirements on the number of jets and the number of $b$-tagged jets.
For the \eightTeV\ dataset three sets of requirements are defined, out of which at least one must be satisfied for an event to be considered:
\begin{enumerate}
\item At least six jets, out of which at least two must be $b$-tagged using the 80\% WP.
\item At least five jets with an increased $\pt$ threshold of 30~\gev, out of which at least two must be $b$-tagged using the 70\% WP.
\item At least six jets with an increased $\pt$ threshold of 30~\gev, out of which at least one must be $b$-tagged using the 60\% WP.
\end{enumerate}
These requirements were optimized to suppress in particular the contribution from \ggH\ Higgs boson production with \hgg\ to the hadronic category,
while retaining good sensitivity to \ttH\ production.
For the \sevenTeV\ dataset only events with at least six jets, at least two of which are $b$-tagged with the 85\% WP, are considered.

Table~\ref{tab:yields} summarizes the expected numbers of events in each category for $\mH = \mHcomb\gev$,
the Higgs boson mass measured by the ATLAS Collaboration~\cite{MassPaper}.
The breakdown into the different Higgs boson production processes is given.
The combined selection efficiencies in the \sevenTeV\ and \eightTeV\ data for \ttH\ production at $\mH = \mHcomb\gev$ are approximately 14.6\% and 14.8\%, respectively.
For SM \tHqb\ (\WtH) production the combined selection efficiencies for \sevenTeV\ and \eightTeV\ are approximately 6.2\% (12.9\%) and 6.2\% (11.9\%), respectively.

\begin{table*}[t]
\centering
\caption{Summary of systematic uncertainties on the final yield of events for \eightTeV\ data from \ttH, \tHqb\ and \WtH\ production after applying the leptonic and hadronic selection requirements.
  The uncertainties are also shown for other Higgs boson production processes that do not include the associated production of top quarks and have significant contributions to the event selection. These are \WH\ production in the leptonic category and \ggH\ production in the hadronic category.
  For both \tH\ production processes, the maximum uncertainty observed for all values of $\kappa_t$ generated (+1, 0, $-1$) is reported.
}
\vspace{0.25cm}
\begin{tabular}{c|cc|cc|cc|c|c}
\multirow{2}{*}{}                  & \multicolumn{2}{c|}{\ttH\ [\%]} & \multicolumn{2}{c|}{\tHqb\ [\%]} & \multicolumn{2}{c|}{\WtH\ [\%]} & \ggH\ [\%] & \WH\ [\%] \\
                                   & had. & lep. & had. & lep. & had. & lep. & had. & lep. \\
\hline
Luminosity                & \multicolumn{8}{c}{$\pm 2.8$} \\
\hline
Photons                   & $\pm 5.6$ & $\pm 5.5$ & $\pm 5.6$ & $\pm 5.5$ & $\pm 5.6$ & $\pm 5.5$ & $\pm 5.6$ & $\pm 5.5$ \\
\hline
Leptons                   & $< 0.1$ & $\pm 0.7$ & $< 0.1$ & $\pm 0.6$ & $< 0.1$ & $\pm 0.6$ & $< 0.1$ & $\pm 0.7$ \\
\hline
Jets and \met             & $\pm 7.4$ & $\pm 0.7$ & $\pm 16$ & $\pm 1.9$ & $\pm 11$ & $\pm 2.1$ & $\pm 29$ & $\pm 10$ \\
\hline
Bkg. modeling             & 0.24 evt. & \multicolumn{1}{c}{0.16 evt.} & \multicolumn{6}{c}{applied on the sum of all Higgs boson production processes} \\
\hline
Theory ($\sigma\times{\rm{BR}}$) & \multicolumn{2}{c|}{+10,$-13$} & \multicolumn{2}{c|}{+7,$-$6} & \multicolumn{2}{c|}{+14,$-$12} & +11,$-$11 & +5.5,$-$5.4 \\
\hline
MC modeling                  & $\pm 11$ & $\pm 3.3$ & $\pm 12$ & $\pm 4.4$ & $\pm 12$ & $\pm 4.6$ & $\pm 130$ & $\pm 100$ \\
\end{tabular}
\label{tab:objectSystematics}
\end{table*}

\section{Analysis}
\label{sec:analysis}

In order to separate processes involving \hgg\ decays from the continuum background, a localized excess of events is searched for 
in the \mgg\ spectrum around $\mH = \mHcomb\gev$.
Probability distribution functions for the \hgg\ resonance and continuum background \mgg\ distributions are defined in the range of 105--160~\gev\ as described below,
and the numbers of Higgs boson and continuum background events are estimated from an unbinned signal-plus-background likelihood fit to the
full \mgg\ distributions in the leptonic and hadronic categories.
Systematic uncertainties are taken into account as nuisance parameters, which are fitted within their external constraints.

The sum of a Crystal Ball function~\cite{crystalball}
and a Gaussian function is used to describe the \mgg\ distribution
from \hgg\ decays obtained from MC simulations~\cite{CouplingPaper}.
The Gaussian function accounts only for a small fraction of the total \hgg\ resonance signal, describing small tails of the shape which cannot be
characterized by the Crystal Ball function.
The parameters of these functions are interpolated between the values fitted to a series of MC samples generated in steps of 5~\gev\ in \mH,
in order to allow for the evaluation of the resonance shape for intermediate masses including $\mH = \mHcomb\gev$, where MC samples are not available.
The relative fraction of the Gaussian component with respect to the full \hgg\ resonance shape is not varied as a function of \mH.
Shapes with different parameter values are defined for the \sevenTeV\ and \eightTeV\ data.
The \mgg\ resolution, which is quantified by 
half of the smallest $\mgg$ interval containing 68\% of the signal events,
is 1.42~\GeV\ for the \sevenTeV\ data and 1.56~\GeV\ for the \eightTeV\ data 
in the leptonic categories. The values in the hadronic categories are consistent with the ones in the leptonic categories within statistical uncertainties. The small difference in \mgg\ resolution between \sevenTeV\ and \eightTeV\ is due to a difference in the effective constant term for the calorimeter energy resolution and due to the lower level of pile-up in the \sevenTeV\ data~\cite{MassPaper}.
The \mgg\ resolution is dominated by the photon energy resolution.
The small change in acceptance for \ttH\ production is
interpolated using MC samples generated with different hypothesized values of \mH\ also.
For all other Higgs boson production processes, the difference in acceptance between $\mH = 125\gev$ and $\mH = \mHcomb\gev$ is found to be negligible.

An exponential function, $\mathrm{e}^{a\,\mgg}$, with $a \leq 0$ is chosen for both categories as a model for the continuum background
following the method previously used in Ref.~\cite{2012gk}.
The choice of fit function is validated in data control regions obtained by loosening the photon identification
and isolation requirements.
These control regions are dominated by jets misidentified as photons, and the systematic uncertainties derived from these control regions 
(cf. Section~\ref{sec:syst}) are hence only approximate.
In both the leptonic and the hadronic category, the same continuum background shape is used for \sevenTeV\ and \eightTeV\ data,
because the \sevenTeV\ data alone is not expected to strongly constrain the parameter $a$ given the expected low number of events.

In the range $105 \gev < \mgg\ < 160\gev$, 3 (3) events are found in the leptonic (hadronic) category in the \sevenTeV\ and
5 (15) events are found in the \eightTeV\ data.
The results of the fits for the leptonic and hadronic categories are shown in Fig.~\ref{fig:fit}, separately for \sevenTeV\ and \eightTeV\ data.
The fitted numbers of continuum background events in a window of 120--130~\gev\ are shown in Table~\ref{tab:yields}.

\section{Systematic uncertainties}
\label{sec:syst}

Systematic uncertainties from various sources affect both the expected number of events for different Higgs boson production processes and the \mgg\ resonance shape.
An overview of all systematic uncertainties for \eightTeV\ data is shown in Table~\ref{tab:objectSystematics} for \ttH, \tHqb\ and \WtH\ production.
The uncertainties are also shown for other Higgs boson production processes that do not include the associated production of top quarks and have significant contributions to the event selection. These are \WH\ production in the leptonic category and \ggH\ production in the hadronic category.

The uncertainty on the integrated luminosity is 2.8\% (1.8\%) for the \eightTeV\ (\sevenTeV) data
as derived following the same methodology as that detailed in Ref.~\cite{lumi} using beam-separation scans.
For \eightTeV\ data, the trigger efficiency~\cite{Aad:2012xs} was measured to be $99.5 \pm 0.2$\%. For \sevenTeV\ data,
the efficiency was measured to be compatible with 100\% within an uncertainty of 0.2\%.
The uncertainty in the combined diphoton identification efficiency is 1.0\% (8.4\%)~\cite{EgammaCalibPaper} for \eightTeV\ (\sevenTeV) data. Due to the high jet multiplicity in this analysis an additional uncertainty of 4\% is added to account for possible mismodeling of the photon identification efficiency.
This additional uncertainty is obtained from data--MC comparisons
of electron efficiencies in $Z(\to ee)$+jets events, where photon identification requirements are applied to the electron clusters~\cite{Aad:2014dya}. 
Analogously, an additional uncertainty of 3\% is assessed for the efficiency of the combined diphoton isolation requirement,
and is added in quadrature to the nominal uncertainty of 2.3\% (2.1\%) in the hadronic (leptonic) category.
The uncertainty on the photon energy scale~\cite{EgammaCalibPaper}
was found to have a negligible effect on the expected yields.
Its effect on the peak position, however, is taken into account, but has a negligible impact on the results.
The uncertainty in the photon energy resolution translates into an uncertainty on the \mgg\ resolution,
and is based on the resolution measured with $Z\rightarrow ee$ events~\cite{EgammaCalibPaper}.
The total \mgg\ resolution uncertainty is 12\% for both the \sevenTeV\ and \eightTeV\ dataset, which is less than 0.2~\GeV.

The uncertainties due to the lepton reconstruction, identification, isolation, and energy/momentum scale and resolution
combine to less than 1\% for all channels.
Uncertainties on the jet energy scale are taken into account, as well as
uncertainties on the jet energy resolution, and on the modeling of the JVF and of the $b$-tagging efficiencies.
All object uncertainties which change the energy or momentum of the corresponding objects are propagated to the \met\ calculation, and
additional uncertainties are taken into account for energy deposits which only enter the
\met\ calculation, but are not part of other objects.

Systematic uncertainties due to the choice of the continuum background fit model are estimated by fitting continuum background distributions
in control regions
with a Higgs boson plus
continuum background model and quantifying the apparent number of Higgs boson events introduced~\cite{2012gk}.
The systematic uncertainty is chosen to be
the maximal apparent number of Higgs boson events in a narrow mass range around \mHcomb~\gev.
Since the contributions from different background processes in the control region may be different from their contributions in the four categories,
the estimate of this uncertainty is approximate, but its impact on the final results is very small.
An uncertainty of 0.24 (0.16) events is estimated in the \eightTeV\ hadronic (leptonic) category as the apparent number of Higgs boson events
under the Higgs boson peak.
For the \sevenTeV\ dataset, uncertainties of 0.12 and 0.01 events are estimated, where all of these numbers have a non-negligible statistical
component from the limited number of events in the control regions considered.
The number of events is lowest in the control region for the hadronic category in 7 TeV data (266 events).

\begin{table*}[b]
\centering
\caption{Observed and expected 95\% CL upper limits on the \ttH\ production cross section times BR(\hgg) relative to the SM cross section times BR(\hgg)
at $\mH = \mHcomb\gev$.
All other Higgs boson production cross sections, including the cross section for \tH\ production, are set to their respective SM expectations.
In addition, the expected limits corresponding to +2$\sigma$, +1$\sigma$, $-$1$\sigma$, and $-$2$\sigma$ variations are shown.
The expected limits are calculated for the case where \ttH\ production is not present.
The results are given for the combination of leptonic and hadronic categories with all systematic uncertainties included, and also for leptonic
and hadronic categories separately, as well as for the expected limits additionally with only statistical uncertainties considered.
}
\vspace{0.25cm}
\begin{tabular}{l|c|c|c|c|c|c}
& Observed limit & Expected limit & +2$\sigma$ & +1$\sigma$ & $-$1$\sigma$ & $-$2$\sigma$\\
\hline
Combined (with systematics)	&6.7	&4.9	&11.9	&7.5	&3.5	&2.6\\
Combined (statistics only)	&	&4.7	&10.5	&7.0	&3.4	&2.5\\
\hline
Leptonic (with systematics)	&10.7	&6.6	&16.5	&10.1	&4.7	&3.5\\
Leptonic (statistics only)	&	&6.4	&15.1	&9.6	&4.6	&3.4\\
\hline
Hadronic (with systematics)	&9.0	&10.1	&25.4	&15.6	&7.3	&5.4\\
Hadronic (statistics only)	&	&9.5	&21.4	&14.1	&6.8	&5.1\\
\end{tabular} 
\label{tab:limit_mu_ttH}
\end{table*}

The theoretical uncertainties on the different Higgs boson production cross sections due to uncertainties in the PDF, missing higher-order perturbative QCD corrections
estimated by varying the renormalization and factorization scales, and the BR(\hgg) are detailed in
Refs.~\cite{Lai:2010vv,Martin:2009iq,Ball:2011mu,Denner:2011mq,LHCHiggsCrossSectionWorkingGroup:2011ti,LHCHiggsCrossSectionWorkingGroup:2012vm,Botje:2011sn,Heinemeyer:2013tqa}.

Additional uncertainties are included in ``MC modeling'' in Table~\ref{tab:objectSystematics}.
These take into account changes in the acceptance when the renormalization and factorization scales are varied, an uncertainty on the modeling of the underlying event,
which is conservatively estimated by comparing MC samples with and without multiple parton scattering,
and an uncertainty due to the limited number of events present in the MC samples after the event selection and categorization are applied.
Moreover, uncertainties of 100\% are assigned to the expected numbers of events from
\ggH, \VBFH\ and \WH\ production in association with additional $b$-jets.
The size of these uncertainties is motivated by recent measurements of $\ttbar$ and vector-boson
production in association with $b$-jets~\cite{Aad:2013tua,Aad:2013vka}. 

\section{Results}

In total, 5 candidate events with \mgg\ in the range 120--130~\gev\ are found in the leptonic and hadronic categories.
The total expected
yield of Higgs boson production is 1.3 events compared to a continuum background of $4.6^{+1.3}_{-0.9}$ events (see Table~\ref{tab:yields}).
The \mgg\ spectra for the candidate events are shown in Fig.~\ref{fig:fit} together with the fitted continuum background
and the total contribution from \hgg\ processes, where the signal strength, \muH, is a parameter common to all four categories.
The best-fit signal strength for all \hgg\ processes together is $1.4^{+2.1}_{-1.4}(\mathrm{stat.})^{+0.6}_{-0.3} (\mathrm{syst.})$, 
where the quoted overall systematic uncertainty is derived by quadratically subtracting the statistical uncertainty from the total uncertainty. 
When the yields for all \hgg\ processes, including \tH\ production but not \ttH\ production, are set to their respective SM expected number of events,
a best-fit value of $1.3^{+2.5}_{-1.7}(\mathrm{stat.})^{+0.8}_{-0.4} (\mathrm{syst.})$ is obtained for \muttH, which
is also shown in the scan of the likelihood in Fig.~\ref{fig:lscan}.
This best-fit value of \muttH\ is consistent with the SM expectation of one, but does not represent a significant excess over the predicted background rate,
and $CL_s$-based~\cite{cls} 95\% CL exclusion upper limits are set
for \ttH\ production times BR(\hgg).
Limits are set using the asymptotic formulae discussed in Ref.~\cite{asym} with the profile likelihood ratio as test statistic.
The results are found to be consistent with limits derived from ensembles of pseudo-experiments.
The observed and expected upper limits for \muttH\ at
$\mH = \mHcomb\gev$ are summarized in Fig.~\ref{fig:overview} as well as in Table~\ref{tab:limit_mu_ttH}, where the expected limits assume $\muttH = 0$. The non-\ttH\ Higgs boson production modes, including \tH, are fixed to their SM expectations with corresponding theory and experimental uncertainties assigned.
An upper limit of 6.7 times the SM cross section times BR(\hgg) is observed.
Upper limits at 95\% CL are also set on the signal strength of the sum of all \hgg\ processes, $\mu$, and the observed (expected) limit is 5.7 (3.8).

\begin{figure}[!h]
  \centering
  \includegraphics[width=0.47\textwidth]{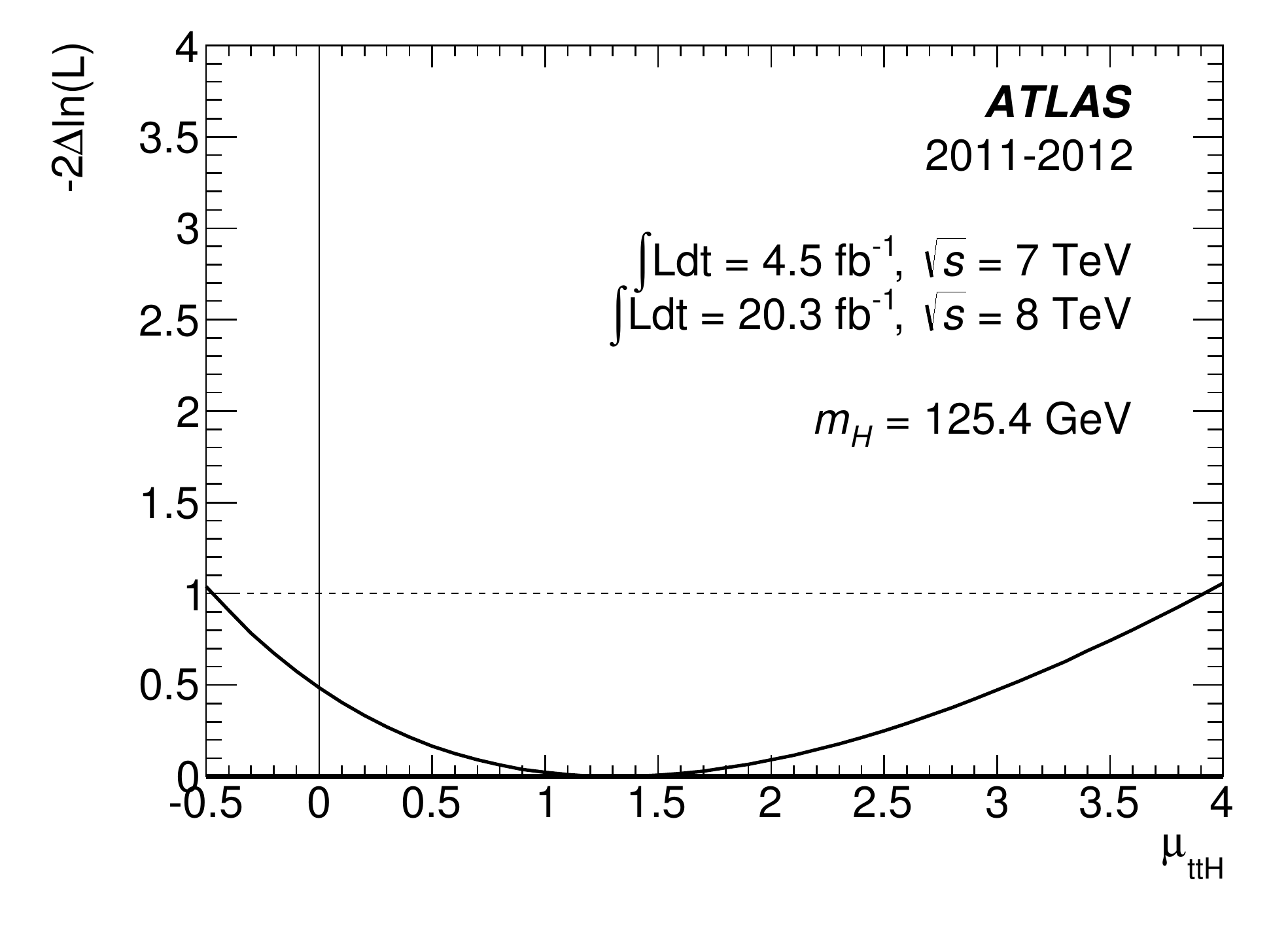}
  \caption{Negative log-likelihood scan for the \ttH\ cross section times BR(\hgg) relative to the SM expectation, \muttH, at $\mH = \mHcomb\gev$,
    where all other Higgs boson production cross sections, including the cross section for \tH\ production, are set to their respective SM expectations.
  }
  \label{fig:lscan}
\end{figure}

\begin{figure}[!h]
  \centering
  \includegraphics[width=0.47\textwidth]{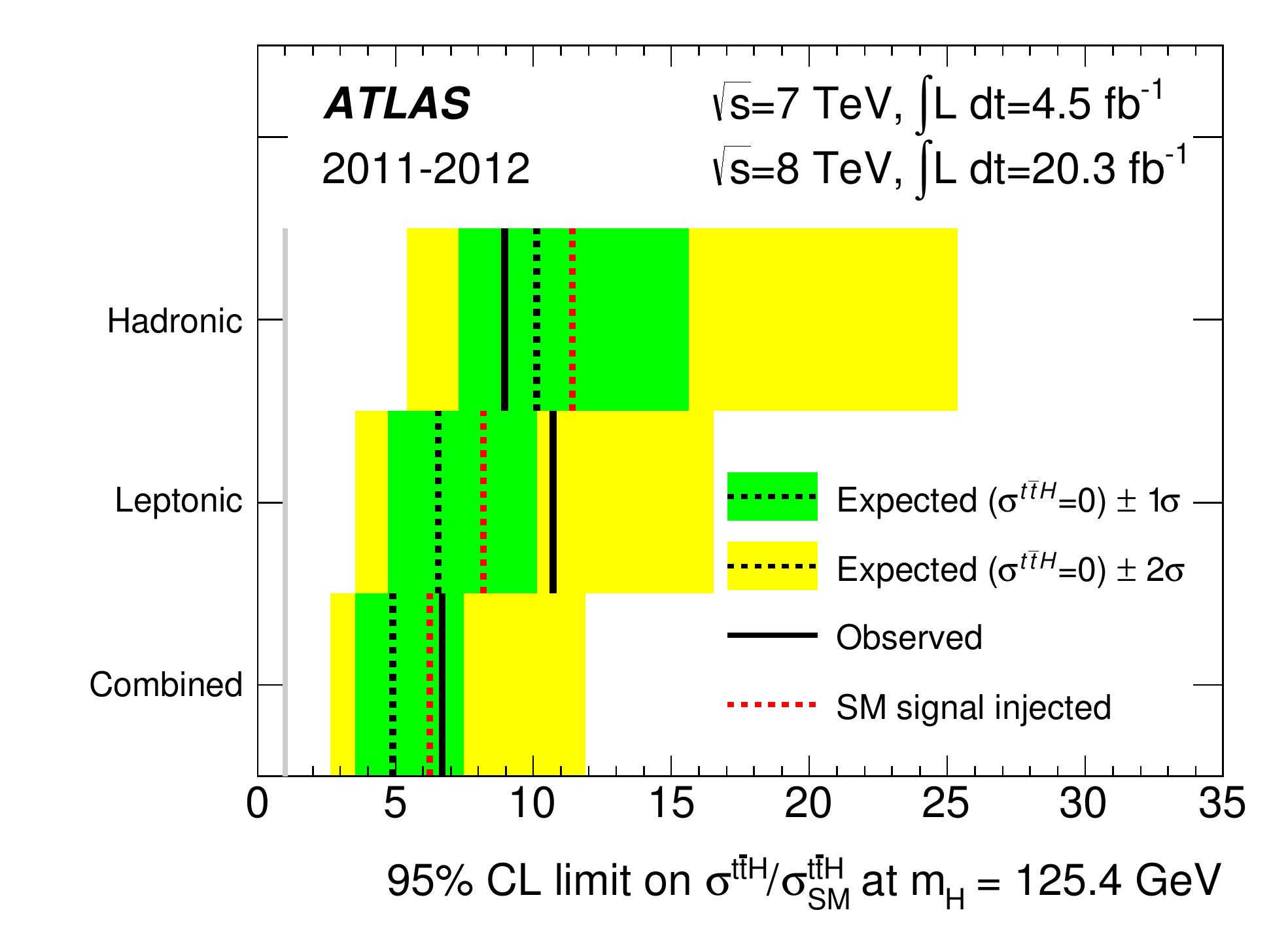}
  \caption{Observed and expected 95\% CL upper limits on the \ttH\ production cross section times BR(\hgg).
    All other Higgs boson production cross sections, including the cross section for \tH\ production, are set to their respective SM expectations.
    While the expected limits are calculated for the case where \ttH\ production is not present,
    the lines denoted by ``SM signal injected'' show the expected 95\% CL limits for a dataset corresponding to continuum background plus SM Higgs boson production.
    The limits are given relative to the SM expectations and at $\mH = \mHcomb\gev$.
  }
  \label{fig:overview}
\end{figure}

\begin{figure}[!h]
  \centering
  \includegraphics[width=0.47\textwidth]{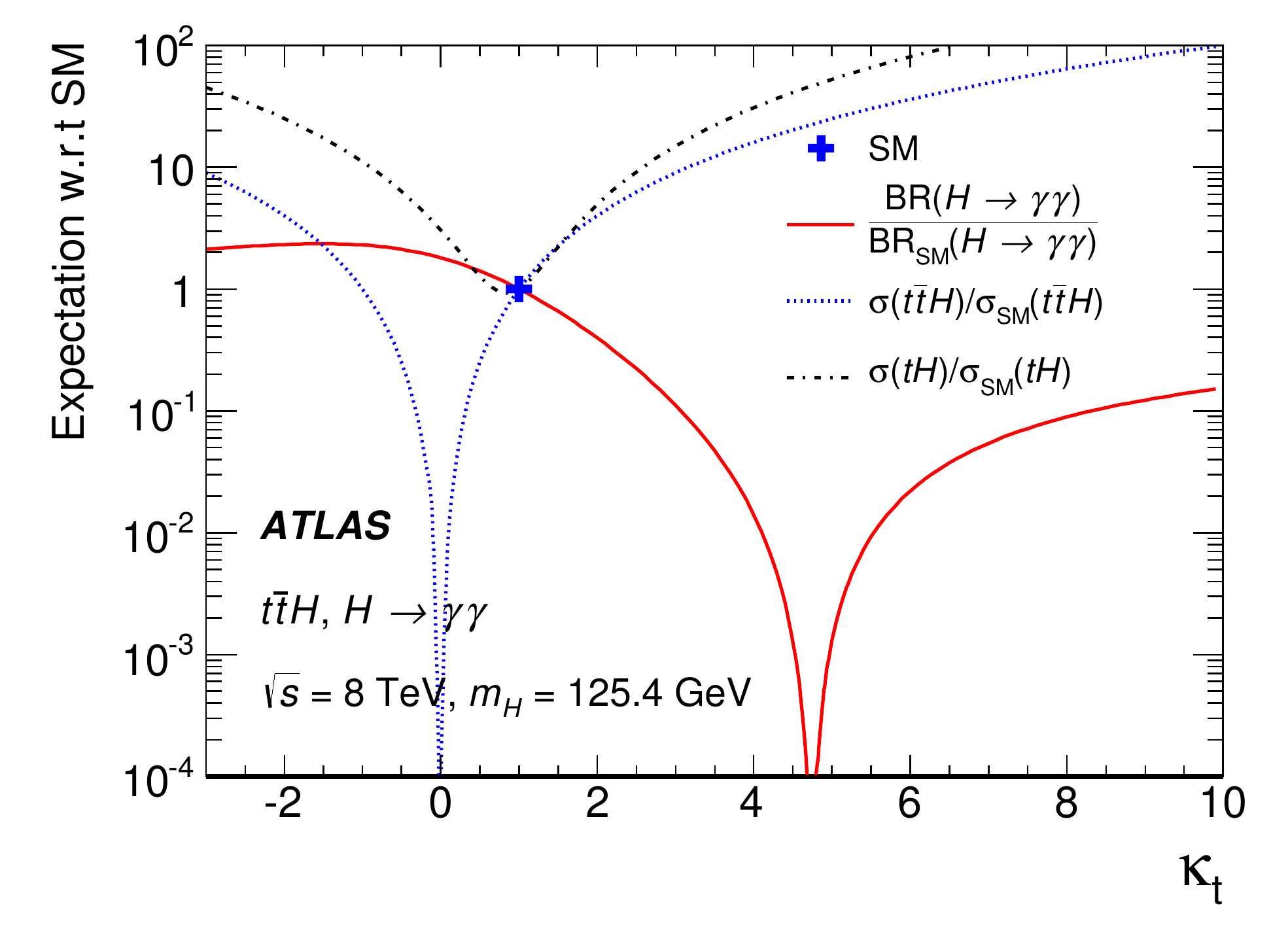}
  \caption{
    Production cross sections for \ttH\ and \tH\ divided by their SM expectations as a function of the scale factor to the top quark-Higgs boson Yukawa coupling, $\kappa_t$.
    Production of \tH\ comprises the \tHqb\ and \WtH\ processes.
    Also shown is the dependence of the BR(\hgg) with respect to its SM expectation on $\kappa_t$.
  }
  \label{fig:properties_Ct}
\end{figure}

\begin{figure}[!h]
  \centering
  \includegraphics[width=0.47\textwidth]{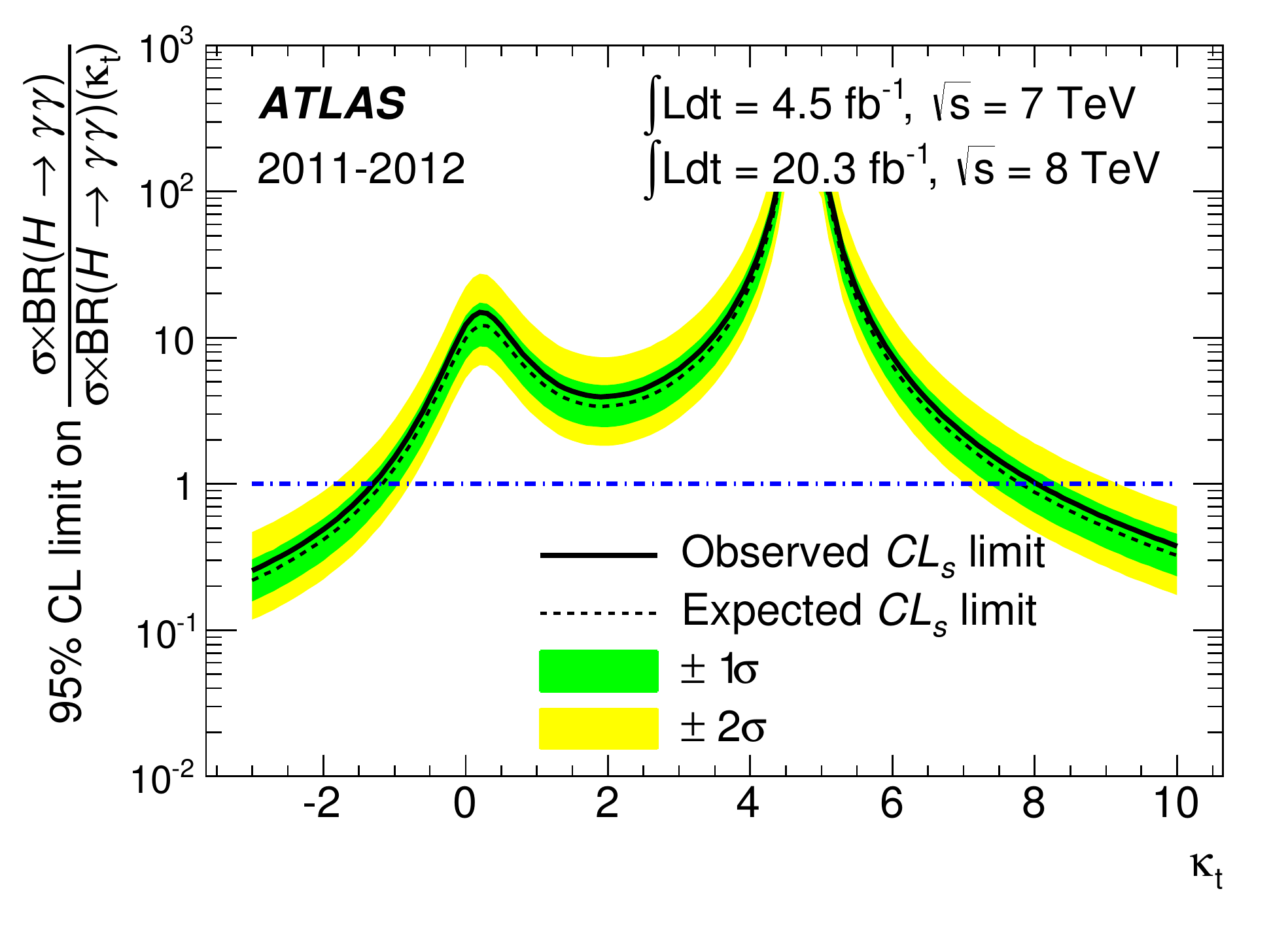}
  \caption{
    Observed and expected 95\% CL upper limits on the inclusive Higgs boson production cross section with respect to the cross section times
    BR(\hgg) for different values of
    $\kappa_t$ at $\mH = \mHcomb\gev$, where $\kappa_t$ is the strength parameter for the top quark-Higgs boson Yukawa coupling.
    All Higgs boson production processes are considered for the inclusive production cross section.
    The expected limits are calculated for the case where $\kappa_t = +1$.
    The $CL_s$ alternative hypothesis is given by continuum background plus SM Higgs boson production.
  }
  \label{fig:overview_Ct}
  \vspace{0.5cm}

  \includegraphics[width=0.47\textwidth]{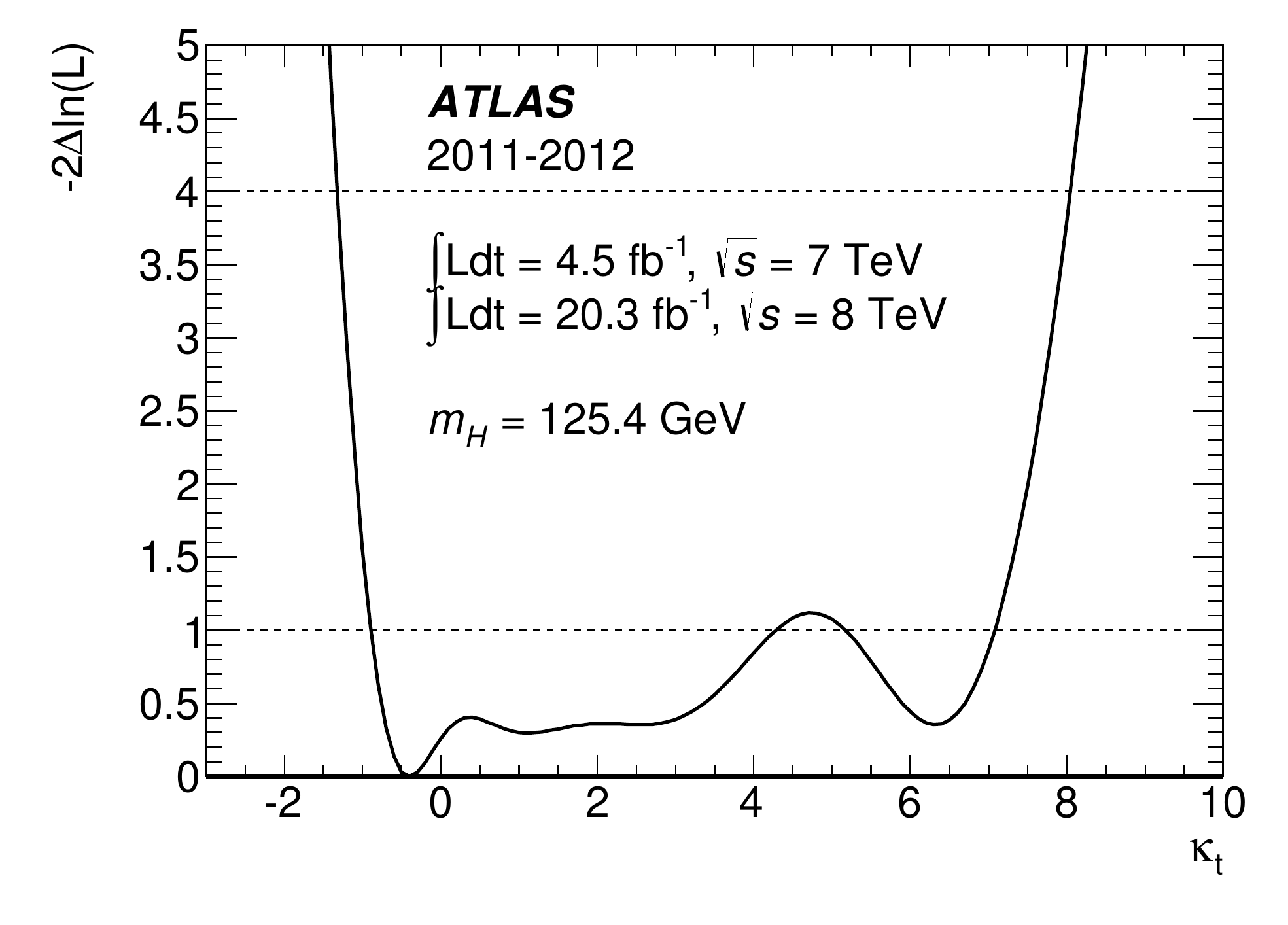}
  \caption{Negative log-likelihood scan of $\kappa_t$ at $\mH = \mHcomb\gev$,
    where $\kappa_t$ is the strength parameter for the top quark--Higgs boson Yukawa coupling.
  }
  \label{fig:nllscan_Ct}
\end{figure}

These results are also interpreted as 95\% CL limits on the strength parameter $\kappa_t$ of the top quark--Higgs boson Yukawa coupling.
Variations in $\kappa_t$ not only change the production cross sections of the \ttH\ and \tH\ processes, but also affect BR(\hgg),
and the cross sections of the other Higgs boson production processes~\cite{Heinemeyer:2013tqa}.
Fig.~\ref{fig:properties_Ct} illustrates the dependence of the \ttH\ and \tH\ cross sections and of the BR(\hgg) on $\kappa_t$.
For $\kappa_t = 0$, the \ttH\ process is turned off, and the top quark contribution to \tH\ production and to the loop-induced \hgg\ decay is removed, leaving mainly the contribution from $W$ bosons. For values of $\kappa_t < 0$, on the other hand, the interference between contributions from $W$ bosons and top quarks to \tH\ production and to the BR(\hgg) becomes constructive, thus enhancing the two processes with respect to their respective SM expectations. Cancellations of the contributions of top quarks and $W$ bosons to the loop-induced \hgg\ decay lead to a minimum of the BR(\hgg) around a value of $\kappa_t = +4.7$.
The combined selection efficiency differs slightly for the three values of $\kappa_t$ for which \tHqb\ and \WtH\ MC samples were generated.
From these, the efficiency at different values of $\kappa_t$ in the range [$-$3,~+10] is calculated by combining reweighted MC samples with $\kappa_t=+1$, $0$ and $-1$. The weight for each sample is assigned in such a way that the cross-section value from the combination follows the prediction shown in Fig.~\ref{fig:properties_Ct}.
The largest relative difference with respect to the efficiency at $\kappa_t = +1$ over the entire range is found to be 14\% (20\%) for \tHqb\ (\WtH) production.

All \hgg\ processes are considered and 95\% CL limits are set on the total Higgs boson production cross section times BR(\hgg) with respect
to the SM cross section for different values of $\kappa_t$.
Coupling strengths other than $\kappa_t$ are set to their respective SM values.
The continuum background plus SM Higgs boson production ($\kappa_t = +1$) is taken as alternative hypothesis.

The observed and expected limits on $\kappa_t$ at $\mH = \mHcomb\gev$ are summarized in Fig.~\ref{fig:overview_Ct}, where the observed (expected)
lower and upper limits on $\kappa_t$ at 95\% CL are $-$1.3 and +8.0 ($-$1.2 and +7.8).
The expected limits assume $\kappa_t = +1$.
The form of the limit curve shown in Fig.~\ref{fig:overview_Ct} is the result of the different dependencies of the different Higgs boson production
processes as well as the BR(\hgg) on $\kappa_t$.
The negative log-likelihood scan of $\kappa_t$ is shown in Fig.~\ref{fig:nllscan_Ct} and it shows that the data are consistent with the SM expectation of $\kappa_t = +1$.
Although two different values of $\kappa_t$ exist with the same total number of expected events, there are no double minima at zero shown in Fig.~\ref{fig:overview_Ct} because different relative contributions from the Higgs boson production processes in different categories have lifted the degeneracy of the likelihood.

\section{Conclusion}

A search for Higgs boson production in association with top quarks in the \hgg\ decay channel is presented using leptonic and hadronic \ttbar\ decays.
Data at \sevenTeV\ and \eightTeV\ corresponding to 4.5~\ifb\ and 20.3~\ifb\ taken in $pp$ collisions with the ATLAS detector at the LHC were analyzed.
No significant excess over the background prediction is observed and upper limits at 95\% CL are set on the \ttH\ production cross section.
The observed exclusion limit at $\mH = \mHcomb\GeV$ is found to be 6.7 times the predicted SM cross section.
The corresponding lower and upper limits on the top quark--Higgs boson Yukawa coupling strength parameter $\kappa_t$ are found to be $-$1.3 and +8.0, which in particular constrain
models with a negative sign of the coupling.


\section{Acknowledgements}

We thank CERN for the very successful operation of the LHC, as well as the
support staff from our institutions without whom ATLAS could not be
operated efficiently.

We acknowledge the support of ANPCyT, Argentina; YerPhI, Armenia; ARC,
Australia; BMWFW and FWF, Austria; ANAS, Azerbaijan; SSTC, Belarus; CNPq and FAPESP,
Brazil; NSERC, NRC and CFI, Canada; CERN; CONICYT, Chile; CAS, MOST and NSFC,
China; COLCIENCIAS, Colombia; MSMT CR, MPO CR and VSC CR, Czech Republic;
DNRF, DNSRC and Lundbeck Foundation, Denmark; EPLANET, ERC and NSRF, European Union;
IN2P3-CNRS, CEA-DSM/IRFU, France; GNSF, Georgia; BMBF, DFG, HGF, MPG and AvH
Foundation, Germany; GSRT and NSRF, Greece; ISF, MINERVA, GIF, I-CORE and Benoziyo Center,
Israel; INFN, Italy; MEXT and JSPS, Japan; CNRST, Morocco; FOM and NWO,
Netherlands; BRF and RCN, Norway; MNiSW and NCN, Poland; GRICES and FCT, Portugal; MNE/IFA, Romania; MES of Russia and ROSATOM, Russian Federation; JINR; MSTD,
Serbia; MSSR, Slovakia; ARRS and MIZ\v{S}, Slovenia; DST/NRF, South Africa;
MINECO, Spain; SRC and Wallenberg Foundation, Sweden; SER, SNSF and Cantons of
Bern and Geneva, Switzerland; NSC, Taiwan; TAEK, Turkey; STFC, the Royal
Society and Leverhulme Trust, United Kingdom; DOE and NSF, United States of
America.

The crucial computing support from all WLCG partners is acknowledged
gratefully, in particular from CERN and the ATLAS Tier-1 facilities at
TRIUMF (Canada), NDGF (Denmark, Norway, Sweden), CC-IN2P3 (France),
KIT/GridKA (Germany), INFN-CNAF (Italy), NL-T1 (Netherlands), PIC (Spain),
ASGC (Taiwan), RAL (UK) and BNL (USA) and in the Tier-2 facilities
worldwide.

\bibliographystyle{elsarticle-num}
\bibliography{paper}

\clearpage
\onecolumn
\begin{flushleft}
{\Large The ATLAS Collaboration}

\bigskip

G.~Aad$^{\rm 84}$,
B.~Abbott$^{\rm 112}$,
J.~Abdallah$^{\rm 152}$,
S.~Abdel~Khalek$^{\rm 116}$,
O.~Abdinov$^{\rm 11}$,
R.~Aben$^{\rm 106}$,
B.~Abi$^{\rm 113}$,
M.~Abolins$^{\rm 89}$,
O.S.~AbouZeid$^{\rm 159}$,
H.~Abramowicz$^{\rm 154}$
% ...
\bigskip
\\
% ...
$^{11}$ Institute of Physics, Azerbaijan Academy of Sciences, Baku, Azerbaijan\\
% ...
$^{84}$ CPPM, Aix-Marseille Universit{\'e} and CNRS/IN2P3, Marseille, France\\
% ...
$^{89}$ Department of Physics and Astronomy, Michigan State University, East Lansing MI, United States of America\\
% ...
$^{106}$ Nikhef National Institute for Subatomic Physics and University of Amsterdam, Amsterdam, Netherlands\\
% ...
$^{112}$ Homer L. Dodge Department of Physics and Astronomy, University of Oklahoma, Norman OK, United States of America\\
$^{113}$ Department of Physics, Oklahoma State University, Stillwater OK, United States of America\\
% ...
$^{116}$ LAL, Universit{\'e} Paris-Sud and CNRS/IN2P3, Orsay, France\\
% ...
$^{152}$ Institute of Physics, Academia Sinica, Taipei, Taiwan\\
% ...
$^{154}$ Raymond and Beverly Sackler School of Physics and Astronomy, Tel Aviv University, Tel Aviv, Israel\\
% ...
$^{159}$ Department of Physics, University of Toronto, Toronto ON, Canada\\
% ...
\end{flushleft}


\end{document}